\documentstyle[12pt]{article}
\def\theequation{\arabic{section}.\arabic{equation}}
\begin{document}
\renewcommand{\thefootnote}{\fnsymbol{footnote}}

\renewcommand{\theequation}{\thesection.\arabic{equation}}

\title{Central extensions, classical non-equivariant maps and residual symmetries}

\author{Francesco Toppan${}^a$\thanks{{\em 
Talk given at the International Conference ``Renormalization Group
and Anomalies in Gravitation and Cosmology", Ouro Preto, Brazil, March 
2003. To appear in the Proceedings.}}
\\ 
 \\ ${}^a${\it CBPF, CCP, Rua Dr.}
{\it Xavier Sigaud 150,}
 \\ {\it cep 22290-180 Rio de Janeiro (RJ), Brazil}\\
 {\it E-mail: toppan@cbpf.br}
}

\maketitle

\begin{abstract}

The arising of central extensions is discussed in two contexts. At first classical 
counterparts  of quantum anomalies (deserving being named as ``classical
anomalies") are associated with a peculiar subclass of the non-equivariant maps. 
Further, the notion of ``residual symmetry" for theories formulated in given 
non-vanishing EM backgrounds is introduced. It is pointed out that this is a 
Lie-algebraic, model-independent, concept.

\end{abstract}

\thispagestyle{empty} \vfill{CBPF-NF-021/03}

\section{Introduction}

We discuss here the contents of two papers, \cite{ft} and \cite{crt},
where two definitions have been proposed for two different phenomenons which are both
related with the arising of centrally extended symmetry algebras. \par
In \cite{ft} the notion of ``classical anomalies" has been introduced to describe a 
classical
counterpart for the well-known notion of quantum anomalies. It can be said that a 
classical anomaly is present
whenever the Noether charges of a given theory, endowed with a classical Poisson 
brackets structure, no longer close the original symmetry algebra of the action, 
but only its
centrally extended version. Classical Poisson brackets are already sufficient to 
produce such an
effect (i.e., it is not necessary to introduce a full commutator algebra for 
quantum operators).
Perhaps the best known example is the Liouville theory \cite{gn}, whose stress-energy tensor,
even classically \cite{ccj}, satisfy $Vir\oplus Vir$, while the original conformal symmetry 
algebra of the classical action is $Witt\oplus Witt$, the direct sum of two copies of 
centerless Virasoro algebras. Even simpler examples can be given \cite{ft}. It is worth 
to point out that a ``classical anomaly" is a very peculiar type of classical non-equivariant 
map.
Indeed, it is a non-equivariant map associated with the Noether charges, i.e. the symmetries, 
of a classical action.\par
The second topic here discussed is the notion of ``residual symmetries"' introduced 
in \cite{crt}.
These ones correspond to the surviving symmetries once an external (for sake of clarity let's 
take an electromagnetic, not necessarily constant) background is turned on.\par
Previous works such as \cite{kar} investigated this issue for very simplified field models 
(e.g., in \cite{kar} a $U(1)$ free massive bosonic field in $1+1$ dimensions, minimally 
coupled to the
external EM background was considered). On the other hand, as shown in \cite{crt} and 
discussed in section $3$,
the notion of ``residual symmetry"  is purely Lie-algebraic and model-independent. 
Any original Lie algebra, or better a D-module realization of it, admits its associated 
residual symmetry.
To give an example, for a generic constant EM background, the Poincar\'e algebra in 
$(2+1)$ dimensions admits as residual symmetry the $5$-generators Lie algebra 
${\cal P}_c(2)\oplus o(2)$,
where ${\cal P}_c (2) $ is the two-dimensional centrally extended Poincar\'e algebra 
discussed
in \cite{cj}. According to the relative strength of the external electric versus 
magnetic field
it could be of Euclidean or Minkowskian type.

\section{Classical anomalies as peculiar non-equivariant maps}

The class of systems under consideration here consists of the
classical dynamical systems which admit both a lagrangian and a
hamiltonian description. It will be further assumed that the
action ${\cal S}$ admits an invariance under a group of symmetries
which can be continuous (Lie), infinite-dimensional and/or super.
The conserved Noether charges are associated to each generator of
the symmetries of the action. When the hamiltonian dynamics is
considered, the phase space of the theory possesses an algebraic
structure given by the Poisson brackets. The existence of such a
structure makes it possible to compute the Poisson bracket between
any two given Noether charges. In the standard situation, the
Poisson brackets among Noether charges realize a closed algebraic
structure which is isomorphic to the original algebra of the
symmetries of the action. It turns out, however, that this is not always
the case. Indeed, it can happen that the algebra of Noether charges
under Poisson bracket structure close a centrally extended version
of the original symmetry algebra. Mimicking the quantum case, the
following definition can be proposed for a classical dynamical
system. The system is said to possess an anomalously realized
symmetry, or in short a ``classical anomaly", if the following
condition is satisfied: the symmetry transformations of the action
admit Noether generators whose Poisson brackets algebra is a
centrally extended version of the algebra of symmetry
transformations.

Therefore a classical anomaly is a very
specific case of ``non-equivariant map" (for a discussion in a
finite-dimensional setting see~\cite{am}). Not all
non-equivariant maps discussed in the literature are classical
anomalies. For instance the one-dimensional free-particle
conserved quantities $p$ (the momentum) and $pt-mx$ generate a
non-equivariant map (the Poisson bracket between $p$ and $pt-mx$
is proportional to the mass $m$). However, despite being
conserved, they do not generate a symmetry of the action and for
that reason they are not Noether charges.

On the other hand,
infinite-dimensional non-equivariant moment maps were con\-struc\-ted
in~\cite{hk}. In those papers the only explicit application
concerned the dynamical systems of KdV type (classical integrable
hierarchies). Such systems, in contrast with the examples
discussed here, admits a hamiltonian description, but not a
lagrangian formulation. Even if conserved quantities can be
constructed, they can not be interpreted as Noether charges.

The possibility for a classical anomaly to occur is based on very
simple and nice mathe\-matical consistency conditions, implemented
by the Jacobi-identity property of the given symmetry algebra. Let
us illustrate this point by considering some generic (but not the
most general) scheme. Let us suppose that the (bosonic) generators
$\delta_a$'s of a symmetry invariance of the action satisfy a
linear algebra whose structure constants satisfy the Jacobi
identity, i.e.
\begin{eqnarray}
[\delta_a,\delta_b] &=&{f_{ab}}^c\delta_c,
\end{eqnarray}
while
\begin{eqnarray}
&[\delta_a,[\delta_b,\delta_c]]
+[\delta_b,[\delta_c,\delta_a]] +[\delta_c,[\delta_a,\delta_b]]=0.&
\end{eqnarray}
The associated Noether charges $Q_a$'s are further assumed to be
the generators of the algebra, i.e., applied on a given field
$\phi$ they produce
\begin{eqnarray}
\delta_a\phi &=& \{Q_a,\phi\}, \label{noter1}
\end{eqnarray}
where the brackets obviously denote the Poisson-brackets.

The condition
\begin{eqnarray}
[\delta_a,\delta_b]\phi &=& {f_{ab}}^c\delta_c \phi,
\label{comm1}
\end{eqnarray}
puts restriction on the possible Poisson brackets algebra
satisfied by the Noether charges. It is certainly true that
\begin{eqnarray}
\{Q_a, Q_b\} &=& {f_{ab}}^c Q_c,
\end{eqnarray}
(which corresponds to the standard case) is consistent with both
(\ref{noter1}) and (\ref{comm1}). However, in a generic case, it
is not at all a necessary condition since more general solutions
can be found. Indeed, the presence of a central extension,
expressed through the relation
\begin{eqnarray}
\{Q_a, Q_b\} &=& {f_{ab}}^c Q_c + k*\Delta_{ab},
\end{eqnarray}
(where $k$ is a central charge and the function $\Delta_{ab}$ is
antisymmetric in the exchange of $a$ and $b$), is allowed.

Indeed, since the relation
\begin{eqnarray}
\{Q_a, \{Q_b,\phi\}\} - \{Q_b, \{Q_a,\phi\}\} &=&
\{\{Q_a,Q_b\},\phi\} \label{pb1}
\end{eqnarray}
holds due to the Jacobi property of the Poisson bracket structure
(which is assumed to be satisfied), no contradiction can be found
with (\ref{comm1}); the right hand side of (\ref{pb1}) in fact is
given by
\begin{eqnarray}
 \{ {f_{ab}}^c Q_c + k*\Delta_{ab}, \phi\} &=& \{
{f_{ab}}^c Q_c, \phi\} = {f_{ab}}^c\delta_c\phi,
\end{eqnarray}
due to the fact that $k$ is a central term and has vanishing
Poisson brackets with any field.

 This observation on one hand puts
restrictions on the possible symmetries for which a classical
anomaly can be detected; the symmetries in question, on a purely
algebraic ground, must admit a central extension. This is not the
case, e.g., for the Lie groups of symmetry based on finite simple
Lie algebras. On the other hand one is warned that, whenever a
symmetry {\it does} admit an algebraically consistent central
extension, it should be carefully checked, for any specific
dynamical model which concretely realizes it, whether it is
satisfied exactly or anomalously. This remark already holds at the
classical level, not just for purely quantum theories.

Some further points deserve to be mentioned. The first one
concerns the fact that the quantization procedure (which, for the
cases we are here considering, can be understood as an explicit
realization of an abstract Poisson brackets algebra as an algebra
of commutators between operators acting on a given Hilbert space)
can induce anomalous terms for theories which, in their classical
version, are not anomalous in the sense previously specified. It
therefore turns out that the occurrence of classical anomalies is
a phenomenon which is ``more difficult to observe" than the
corresponding appearance of quantum anomalies since it occurs more
seldom.

A second point concerns the fact that the algebra of
Poisson brackets, as an abstract algebra, is assumed to satisfy
the Leibniz property. This is no longer the case for its concrete
realization given by the algebra of commutators. The Noether
charges are in general non-linearly constructed with the original
fields $\phi_i$ (which collectively denote the basic fields and
their conjugate momenta) of a given theory. For such a reason it
is only true in the classical case that, whenever an anomalous
central charge in an infinitesimal linear algebra of symmetries is
detected, it can be normalized at will by a simultaneous rescaling
of all the fields $\phi_i$ involved
($\phi_i\mapsto\alpha\cdot\phi_i$) and of the Poisson brackets
normalization ($\{.,.\}\mapsto \frac{1}{\alpha}
\{.,.\}$), for an arbitrary real constant $\alpha$. In the
classical case any central charge different from zero can
therefore be consistently set equal to $1$. However in the quantum
case a specific value of the central charge is fixed by the type
of representation of the symmetry algebra associated with the
given model and is a genuine physical parameter (the role of the
Virasoro central charge in labeling the conformal minimal models
is an example). The above argument is not, however, (at least
directly) applicable to non-linear symmetries, such as those
leading to the classical counterparts of the Fateev--Zamolodchikov
$W$-algebras. 

\section{Residual symmetries in the presence of an
EM background}

Let us discuss in detail for the sake of simplicity the case of
the residual symmetry for generic Poincar\'e-invariant field
theories in $(2+1)$-dimension, coupled with an external constant
EM background. The generalization of this procedure to
higher-dimensional theories and non-constant EM backgrounds is
straightforward and immediate.\par In the absence of the external
electric and magnetic field, the action ${\cal S}$ is assumed to
be invariant under a $7$-parameter symmetry, given by the six
generators of the $(2+1)$-Poincar\'e symmetry which, when acting
on scalar fields (the following discussion however is valid no
matter which is the spin of the fields) are represented by
\begin{eqnarray}
P_\mu &=& -i\partial_\mu, \nonumber\\ M_{\mu\nu} &=& i(x_\mu
\partial_\nu -x_\nu\partial_\mu),\label{poinc}
\end{eqnarray}
(the metric is chosen to be $+--$), plus a remaining symmetry
generator corresponding to the internal global $U(1)$ charge that
will be denoted as $Z$.
\par
It is further assumed that in the action ${\cal S}$ the dependence
on the classical background field is expressed in terms of the
covariant gauge-derivatives \begin{eqnarray} { D}_\mu &=&
\partial_\mu - ieA_\mu, \nonumber \end{eqnarray} with $e$ the
electric charge.\par
\par
In the presence of constant external electric and magnetic fields,
the $F^{\mu\nu}=\partial^\mu A^\nu-\partial^\nu A^\mu$
field-strength is constrained to satisfy
\begin{eqnarray}
F^{0i} = E^i, &\quad F^{ij} = \epsilon^{ij} B,\label{const}
\end{eqnarray}
where $\mu,\nu =0,1,2$ and $i,j=1,2$.  The fields $E^i$ and $B$
are constant. Without loss of generality the $x^1$, $x^2$ spatial
axis can be rotated so that $E^1\equiv E$, $E^2=0$. Throughout the
text this convention is respected.\par In order to recover
(\ref{const}), the gauge field $A_\mu$ must depend at most
linearly on the coordinates $x^0\equiv t$, $x^1\equiv x$ and
$x^2\equiv y$.\par The gauge-transformation
\begin{eqnarray}
&&A_\mu \mapsto {A_\mu}' = A_\mu +\frac{1}{e}\partial_\mu
\alpha(x^\nu) \label{gtrans}
\end{eqnarray}
allows to conveniently choose for $A_\mu$ the gauge-fixing
\begin{eqnarray}
A_0 &=& 0,\nonumber\\ A_i &=& E_i t -\frac{B}{2}\epsilon_{ij} x^j.
\label{gfix}
\end{eqnarray}
The above choice is a good gauge-fixing since it completely fixes
the gauge (no gauge-freedom is left). It will be soon evident that
the residual symmetry is a truly physical symmetry, independent of
the chosen gauge-fixing.\par Due to (\ref{gfix}), the action
${\cal S}$ explicitly depends on the $x^\mu$ coordinates entering
$A_\mu$. The simplest way to compute the symmetry property of an
action such as ${\cal S}$ which explicitly depends on the
coordinates consists in performing the following trick. At first
$A_\mu$ is regarded on the same foot as the other fields entering
${\cal S}$ and assumed to transform as standard vector field under
the global Poincar\'e transformations, namely
\begin{eqnarray}
{A_\mu}' ({x^\rho}') &=& {\Lambda_\mu}^\nu A_\nu (x^\rho)
\end{eqnarray}
for $ {x^\mu}' = {\Lambda^\mu}_\nu x^\nu + a^\mu $.\par For a
generic infinitesimal Poincar\'e transformation, however, the
transformed $A_\mu$ gauge-field no longer respects the
gauge-fixing condition (\ref{gfix}). In the active transformation
viewpoint only fields are entitled to transform, not the
space-time coordinates themselves. $A_\mu$ plays the role of a
fictitious field, inserted to take into account the dependence of
the action ${\cal S}$ on the space-time coordinates caused by the
non-trivial background. Therefore, the overall infinitesimal
transformation $\delta A_\mu$ should be vanishing. This result can
be reached if an infinitesimal gauge transformation (\ref{gtrans})
$\delta_g(A_\mu)$ can be found in order to compensate for the
infinitesimal Poincar\'e transformation $\delta_P(A_\mu)$, i.e. if
the following condition is satisfied
\begin{eqnarray}
&&\delta(A_\mu) =\delta_P(A_\mu) +\delta_{g}(A_\mu)
=0.\label{delta}
\end{eqnarray}
Only those Poincar\'e generators which admit a compensating
gauge-transformation satisfying (\ref{delta}) provide a symmetry
of the ${\cal S}$ action (and therefore enter the residual
symmetry algebra). This is a plain consequence of the original
assumption of the Poincar\'e and manifest gauge invariance for the
action ${\cal S}$ coupled to the gauge-field $A_\mu$.\par Notice
that the original Poincar\'e generators are deformed by the
presence of extra-terms associated to the compensating gauge
transformation. Let $p$ denote a generator of (\ref{poinc}) which
``survives" as a symmetry in the presence of the external
background. The effective generator of the residual symmetry is
\begin{eqnarray}
{\hat p} &=& p+(\ldots),\nonumber \end{eqnarray} where $(\ldots )$
denotes the extra terms arising from the compensating gauge
transformation associated to $p$. Such $(\ldots )$ extra terms are
gauge-fixing dependent. The ``residual symmetry generator" ${\hat
p}$ can only be expressed in a gauge-dependent manner. However,
two gauge-fixing choices are related by a gauge transformation
${\bf g}$. The residual symmetry generator in the new
gauge-fixing, denoted as ${\tilde p}$, is related to the previous
one by an Adjoint transformation
\begin{eqnarray}
{\tilde p}&=& {\bf g} {\hat p} {\bf g}^{-1}. \end{eqnarray}
Therefore the residual symmetry algebra does not dependent on the
choice of the gauge fixing and is a truly physical
characterization of the action ${\cal S}$.\par The extra-terms
$(\ldots )$ are necessarily linear in the space-time coordinates
when associated with a translation generator, and bilinear when
associated to a surviving Lorentz generator (for a constant EM
background). Their presence implies the arising of the central
term in the commutator of the deformed translation generators.

The residual symmetry algebra of the $(2+1)$-Poincar\'e theory
involves, besides the global $U(1)$ generator $Z$, the three
deformed translations and just one deformed Lorentz generator (the
remaining two Lorentz generators are broken).\par Within the
(\ref{gfix}) gauge-fixing choice the deformed translations are
explicitly given by
\begin{eqnarray}
P_0 &=& -i\partial_t -e E x,\nonumber\\ P_1 &=&
-i\partial_x-\frac{e}{2}B y,\nonumber\\ P_2 &=& -i\partial_y
+\frac{e}{2}B x.\label{trasl}
\end{eqnarray}
The deformed generator of the residual Lorentz symmetry is
explicitly given, in the same gauge-fixing and for $E\neq 0$, by
\begin{eqnarray}
M &=& i(x\partial_t + t\partial_x) - i\frac{B}{E}
(y\partial_x-x\partial_y) +\nonumber\\&&\frac{e}{2}(E
t^2+Ex^2-Bty).\label{Lor}
\end{eqnarray}
The residual symmetry algebra can be easily computed.
The $U(1)$ charge $Z$ is no longer decoupled from the other
symmetry generators. It appears instead as a
central charge.\par 
\par The $5$-generator solvable, non-simple Lie algebra of
residual symmetries admits a convenient presentation. The
generator
\begin{eqnarray} {\tilde Z}\equiv B P_0 + E P_2
\label{tildez}\end{eqnarray}
not only commutes with all the other $\ast$ generators
\begin{eqnarray}
\relax [ {\tilde Z}, \ast ] &=&0,
\end{eqnarray}
for $E\neq B$ it is not even present in the r.h.s.,
so that the residual symmetry algebra is given by a direct sum of
$u(1)$ and a $4$-generator algebra. The latter algebra is
isomorphic to the centrally extended two-dimensional Poincar\'e
algebra. Such an algebra is of Minkowskian or Euclidean type
according to whether $ E> B$ or respectively $E<B$. 
This point can be intuitively understood due to
the predominance of the electric or magnetic effect (in the
absence of the electric field the theory is manifestly rotational
invariant, so that the Lorentz generator is associated with the
Euclidean symmetry). We have explicitly, for $B>E$, that the
algebra
\begin{eqnarray}
\relax [{\overline M}, S_1]&=& i S_2,\nonumber\\ \relax
[{\overline M}, S_2] &=& -i S_1
\end{eqnarray}
is reproduced by
\begin{eqnarray}
{\overline M} &=&  \frac{E}{\sqrt{B^2-E^2}}M,\nonumber\\ S_1 &=&
P_0 + \frac{B}{E} P_2,\nonumber\\ S_2 &=& \frac{\sqrt{B^2-E^2}}{E}
P_1,
\end{eqnarray}
while for $E>B$ the algebra
\begin{eqnarray}
\relax [{\tilde M}, T_1]&=& i T_2,\nonumber\\ \relax [{\tilde M},
T_2] &=& i T_1,
\end{eqnarray}
is reproduced by
\begin{eqnarray}
{\tilde M}&=&\frac{E}{  \sqrt{E^2-B^2}} M,\nonumber\\ T_1 &=& P_0
+ \frac{B}{E} P_2,\nonumber\\ T_2 &=& -\frac{\sqrt{E^2-B^2}}{E}
P_1.
\end{eqnarray}
In both cases the commutator between the translation generators
$S_1$, $S_2$, and respectively $T_1$, $T_2$, develops the central
term proportional to $Z$ which can be conveniently normalized.\par
The residual symmetry algebra of the $(2+1)$ case for generic
values of $E$ and $B$ (the special case $E=B$ is degenerate) 
is therefore
given by the direct sum \begin{eqnarray} &&u(1)\oplus {\cal P}_c
(2).\end{eqnarray}

\end{document}